\documentclass[conference,a4paper]{IEEEtran}

\usepackage{flushend}
\usepackage[utf8]{inputenc} 
\usepackage[T1]{fontenc}
\usepackage{url}              
\usepackage{cite}             

\usepackage[cmex10]{amsmath}  
\usepackage{mleftright}       
\mleftright                   

\usepackage{graphicx}         
\usepackage{booktabs}         

\usepackage{times,amsmath,amsfonts,units,fullpage,mathtools}
\usepackage{subcaption}

\newcommand{\E}{\mathbb E}

\usepackage{color}

\newtheorem{theorem}{Theorem}

\newtheorem{definition}{Definition}
\makeatletter
\DeclarePairedDelimiter\p{\lparen}{\rparen}
\let\oldp\p
\def\p{\@ifstar{\oldp}{\oldp*}}

\DeclarePairedDelimiter\bracket{\lbrack}{\rbrack}
\def\bracket{\@ifstar{\oldbracket}{\oldbracket*}}

\DeclarePairedDelimiter\set\{\}
\let\oldset\set
\def\set{\@ifstar{\oldset}{\oldset*}}

\DeclarePairedDelimiter\abs{\lvert}{\rvert}
\let\oldabs\abs
\def\abs{\@ifstar{\oldabs}{\oldabs*}}

\DeclarePairedDelimiter\norm{\lVert}{\rVert}
\let\oldnorm\norm
\def\norm{\@ifstar{\oldnorm}{\oldnorm*}}

\DeclarePairedDelimiter{\ib}{\llbracket}{\rrbracket}
\let\oldib\ib
\def\ib{\@ifstar{\oldib}{\oldib*}}
\makeatother

\DeclareMathOperator{\Diag}{Diag}
\DeclareMathOperator{\Unif}{Unif}

\begin{document}
\title{High-Dimensional Smoothed Entropy Estimation via Dimensionality Reduction}
\author{%
}
%
%
 \author{%
   \IEEEauthorblockN{Kristjan Greenewald}
   \IEEEauthorblockA{IBM Research\\
                     MIT-IBM Watson AI Lab\\
                     kristjan.h.greenewald@ibm.com}
   \and
   \IEEEauthorblockN{Brian Kingsbury}
   \IEEEauthorblockA{IBM Research\\
                     MIT-IBM Watson AI Lab\\
                     bedk@ibm.com}
   \and
   \IEEEauthorblockN{Yuancheng Yu}
   \IEEEauthorblockA{UIUC\\
                     yyu51@illinois.edu}
 }
%
%

\maketitle

\begin{abstract}
We study the problem of overcoming exponential sample complexity in differential entropy estimation under Gaussian convolutions. Specifically, we consider the estimation of the differential entropy $h(X+Z)$ via $n$ independently and identically distributed samples of $X$, where $X$ and $Z$ are independent $D$-dimensional random variables with $X$ sub-Gaussian with bounded second moment and $Z\sim\mathcal{N}(0,\sigma^2I_D)$. Under the absolute-error loss, the above problem has a parametric estimation rate of $\frac{c^D}{\sqrt{n}}$, which is exponential in data dimension $D$ and often problematic for applications. We overcome this exponential sample complexity by projecting $X$ to a low-dimensional space via principal component analysis (PCA) before the entropy estimation, and show that the asymptotic error overhead vanishes as the unexplained variance of the PCA vanishes. This implies near-optimal performance for inherently low-dimensional structures embedded in high-dimensional spaces, including hidden-layer outputs of deep neural networks (DNN), which can be used to estimate mutual information (MI) in DNNs. We provide numerical results verifying the performance of our PCA approach on Gaussian and spiral data. We also apply our method to analysis of information flow through neural network layers (c.f. information bottleneck), with results measuring mutual information in a noisy fully connected network and a noisy convolutional neural network (CNN) for MNIST classification.\footnote{The appendices referenced in this paper along with additional experiments are hosted on arXiv at \url{https://arxiv.org/abs/2305.04712}.}\footnote{The authors would like to thank Ziv Goldfeld for helpful discussions during the formation of the project.}
\end{abstract}

\section{Introduction}
This work addresses the problem of the estimation of smoothed entropy (and associated measures) in high dimensions, also known as ``differential entropy estimation under Gaussian convolutions.'' For a given random vector $X \in \mathbb{R}^D$, the ($\sigma$-) smoothed entropy was introduced in \cite{goldfeld2019estimating,goldfeld2020convergence} as 
\[
h_\sigma(X) = h(X + Z), \quad Z \sim \mathcal{N}(0,\sigma^2 I_D)
\]
where $\mathcal{N}(0,\sigma^2 I_D)$ is isotropic Gaussian noise with variance $\sigma^2$.\footnote{While we retain isotropic noise for simplicity of presentation, it can be extended to non-isotropic noise via normalization.} Our goal is to estimate this quantity based on samples of $X$, when the dimensionality $D$ is large. This nonparametric functional estimation problem generally requires a number of samples exponential in dimension \cite{goldfeld2020convergence}, in this work we propose using dimensionality reduction to address this issue for a class of approximately low-dimensional distributions.

A key application of smoothed entropy is in estimation of mutual information in noisy channels, 
\begin{equation}\label{eq:smoothMI}
\mathsf{I}(X; Y + Z) = h(Y+Z) - h(Y+Z | X),
\end{equation}
where we assume $Y \in \mathbb{R}^D$, $Z\sim \mathcal{N}(0,\sigma^2 I_D)$ is Gaussian noise and the conditional distribution $Y | X=x \sim p_{Y|X=x}$ can be sampled from. In this case, $X \in \mathbb{R}^{D_X}$ where $D_X$ does not directly affect the scaling of the estimator. An example application of this framework to machine learning is the analysis of information flow in noisy neural networks \cite{goldfeld2019estimating}.

It can also be used in the joint sampling case, where noise is added to both arguments. This corresponds to the mutual information between outputs of a pair of noisy channels with coupled inputs.  Specifically, 
\begin{align}
\mathsf{I}(X + Z_1, Y + Z_2) =& h(X + Z_1) + h(Y + Z_2) \nonumber\\&- h(X + Z_1, Y + Z_2),\label{eq:smoothjointMI}
\end{align}
where here we assume equal dimensions $X, Y \in \mathbb{R}^{D}$ and equal noise levels $Z_1, Z_2 \sim \mathcal{N}(0, \sigma^2 I_D)$.\footnote{The extension to non-equal dimensions is clear, and the extension to non-equal or non-isotropic noise is immediate via normalization.} Note all terms of \eqref{eq:smoothjointMI} are smoothed entropies.

It was noted in \cite{goldfeld2019estimating} that smoothed entropies cannot be computed in closed form.
The work in \cite{goldfeld2019estimating,goldfeld2020convergence} presented a plug-in estimator for smoothed entropy, and established rates of convergence of the form $O(c^D n^{-1/2})$, where $n$ is the number of samples. While this rate converges quickly with $n$,\footnote{Compare to the $O(n^{-1/D})$ rate of non-smoothed entropy, which is a key motivation for using smoothed entropy \cite{goldfeld2020convergence}} the prefix grows exponentially with dimension $D$ and thus quickly becomes impractical for high dimensional settings.

High dimensional data is ubiquitous in statistics and machine learning, and being able to estimate entropy in these regimes and apply information-theoretic quantities is highly desirable. Example applications include the study of information flow in neural networks \cite{tishby_DNN2015,goldfeld2019estimating,xu2017information,steinke2020reasoning}, independence testing \cite{berrett2019nonparametric}, conditional independence testing for causal structure learning \cite{sen2017model}, regularization of neural network architectures \cite{belghazi2018mine}, generative modeling (e.g. InfoGAN \cite{chen2016infogan}), and many others.

The application of estimating information flows in DNNs is particularly illustrative. There has been a surge of interest in measuring the mutual information between selected groups of neurons in a DNN \cite{DNNs_ICLR2018,jacobsen2018revnet,liu2018understanding,gabrie2018entropy,reeves2017additivity,goldfeld2019estimating}, partially driven by the Information Bottleneck (IB) theory \cite{tishby_DNN2015,DNNs_Tishby2017}. Typically, focus centers on the mutual information $\mathsf{I}(X;T)$ between the input feature $X$ and the values of a hidden neural network layer $T$. However, as explained in \cite{goldfeld2019estimating}, this quantity is vacuous in deterministic DNNs\footnote{i.e., DNNs that, for fixed parameters, define a deterministic mapping from input to output.} and becomes meaningful only when a mechanism for discarding information (e.g., noise) is integrated into the system. To remedy this, a noisy DNN framework was proposed in \cite{goldfeld2019estimating}, where each neuron adds a small amount of Gaussian noise (i.i.d. across neurons) after applying the activation function. In the noisy DNN framework of \cite{goldfeld2019estimating} and \cite{goldfeld2020convergence}, each neuron adds $\sigma$-Gaussian noise to its output. Hence, $\mathsf{I}(X;T)$ becomes non-vacuous:
\begin{equation*}
\mathsf{I}(X;T) = \mathsf{I}(X;T' + Z) = h(T' + Z) - h(T' + Z | X),
\end{equation*}
where $T'$ is the output of the neuron prior to added noise, and $Z \sim \mathcal{N}(0, \sigma^2 I)$. Note that this reduces the mutual information of interest to the smoothed differential entropies we are considering in the present work. The task then reduces to estimating these smoothed entropies from samples obtained from propagating available samples of $X$ through the noisy DNN. Unfortunately, however, the dimensionality of $T$ is often large, and \cite{goldfeld2019estimating} identified significant issues in scaling the estimator to non-toy neural networks. 

To remedy this curse of dimensionality ($c^D$ scaling), we propose to estimate smoothed entropy in high dimensions by first \emph{projecting the data to a lower $d$-dimensional space} prior to estimating the smoothed entropy. This is motivated by the \emph{manifold hypothesis of machine learning} \cite{bengio2013representation}, which states that most high-dimensional data distributions have support on or near a much lower dimensional \emph{data manifold}. This data manifold may be linear or near-linear (in which case it is a subspace discoverable by Principal Component Analysis), or highly nonlinear. 

Smoothed entropy is particularly well suited to the approximately-low-dimensional regime. While projecting away low-variance dimensions can have an unbounded (even infinite) impact on traditional entropy, since smoothed entropy involves convolution with a $\sigma$-Gaussian, the entropy lost in the projection operation can be lower bounded in terms of $\sigma$ and upper bounded in terms of the variance of the removed dimensions. 
Using this fact, we provide a theoretical foundation for performing dimensionality reduction prior to smoothed entropy information estimation, without the need for smoothness assumptions on the distribution of $X$. This enables principled use of powerful information-theoretic measures in high dimensional regimes.

In passing, we point out that without smoothing, entropy (and mutual information) estimation in high dimensions scales even more poorly with dimension than the smoothed case. In particular, the minimax convergence rate for any estimator is on the order of $n^{-1/D}$, with a constant multiplier that may also grow with dimension \cite{jiao2017nearest,Hero_EDGE2018}. While practical success has been seen with neural estimators (e.g. MINE \cite{belghazi2018mine}), it has been shown that these approaches cannot fully defeat the curse of dimensionality \cite{poole2019variational}. Alternatively, averaging the mutual information over random projections to lower dimensional space was proposed as Sliced Mutual Information \cite{goldfeld2022k} -- while the estimation of Sliced Mutual Information does not directly suffer from a curse of dimensionality, it is a separate quantity from traditional mutual information and does not retain all properties thereof, e.g. the data processing inequality. Our work differs in that we attempt to estimate the classical mutual information directly. 

Here for space reasons we focus on Principal Component Analysis (PCA) dimensionality reduction. Our approach, however, can be in principle be generalized to arbitrary, possibly nonlinear, dimensionality reduction approaches (e.g. isomap \cite{balasubramanian2002isomap} or self-supervised neural network embeddings such as in \cite{zbontar2021barlow}).



\section{Smoothed Entropy Estimation via Dimensionality Reduction}

We first establish some notation. Recall $X \in \mathbb{R}^D$ is a random vector with density $p_X$. Suppose we have $2n$ samples $\{X_i\}_{i=1}^{2n}$ from $p_X$. 

We aggregate these samples into two sample matrices of (for simplicity) equal size:
\begin{align*}
\mathbf{X}^{(1)}_n = \left[ \begin{array}{ccc} X_1 & \cdots & X_n\end{array}\right],\\
\mathbf{X}^{(2)}_n = \left[ \begin{array}{ccc} X_{n+1} & \cdots & X_{2n}\end{array}\right],
\end{align*}
where the samples $\mathbf{X}^{(1)}_n$ are used to estimate the projection operator, and the independent samples $\mathbf{X}^{(2)}_n$ are used to estimate the entropy of the resulting projected copy of $X$. This splitting of samples ensures independence of the projection and the samples used to estimate the projected entropy, which simplifies the derivation of the theoretical bounds. 
Let the covariance matrix $\Sigma = \mathbb{E}[X X^T]$ and the sample covariance
\[
\Sigma_n = \frac{1}{n} \sum_{i=1}^n X_i X_i^T.
\]
Since $\Sigma$, $\Sigma_n$ are positive semidefinite by definition, we can denote the eigenvalue decompositions of $\Sigma$, $\Sigma_n$ as
\[
\Sigma = \sum_{j = 1}^D \lambda_j V_j V_j^T, \quad \Sigma_n = \sum_{j = 1}^D \hat\lambda_j \hat V_j \hat V_j^T
\]
respectively, where $\lambda_j$, $\hat{\lambda}_j$ are the corresponding eigenvalues sorted in descending order\footnote{$\lambda_j$, $\hat{\lambda}_j$ are real and $\geq 0$ since the matrices are pos. semidefinite.} and $V_j ,\hat{V}_j \in \mathbb{R}^D$ are the corresponding (orthonormal) eigenvectors.

By using PCA, we implicitly assume (to be made more formal in the next section) that $d$ is chosen such that the top $d$ eigenvalues dominate and the $d+1$th eigenvalue and following are small (i.e. $\lambda_{d+1}$ is small). 

We can then define PCA dimensionality reduction, which corresponds to the projection of the data $X$ onto the top $d$ eigenvectors of $\Sigma_n$. Specifically, let 
\[
\mathbf{V}_d = \left[ \begin{array}{ccc} V_1 & \cdots & V_d\end{array}\right], \quad \hat{\mathbf{V}}_d = \left[ \begin{array}{ccc} \hat V_1 & \cdots & \hat V_d\end{array}\right]
\]
be the $D \times d$ matrices of the top $d$ eigenvectors of $\Sigma$ and $\Sigma_n$ respectively. Letting $\Pi_{\mathbf{V}}$ denote the projection operator onto the span of $\mathbf{V}$, define
\[
X_{\mathbf{V}_d} = \Pi_{\mathbf{V}_d} X = \mathbf{V}_d^T X,\quad X_{\hat{\mathbf{V}}_d} = \Pi_{\hat{\mathbf{V}}_d} X = \hat{\mathbf{V}}_d^T X.
\]
$X_{\hat{\mathbf{V}}_d}$ is the empirical PCA dimensionality reduction to $d$ dimensions using the sample covariance $\Sigma_n$, and $X_{\mathbf{V}_d}$ is the corresponding dimensionality reduction using the true covariance $\Sigma$.

Our goal is to approximate the entropy of $X$ by the entropy of the dimensionality-reduced $X_{\hat{\mathbf{V}}_d} = \Pi_{\hat{\mathbf{V}}_d} X$, and estimate this dimension-reduced entropy using the samples contained in $\mathbf{X}^{(2)}_n$. Since the projected copy has lower dimension, such an approximation requires an additive correction term to account for the entropy in the deleted dimensions. As we are assuming that $X$ has low variance/entropy in the deleted dimensions, for projections onto linear subspaces (e.g. PCA) we can lower bound the entropy in the $D-d$ deleted dimensions by the entropy of $D-d$ dimensional $\sigma^2$-variance Gaussian noise. In the case of the oracle projection $\Pi_{\mathbf{V}_d} X$, we can also upper bound it by the entropy of a Gaussian with covariance matching that of $X$ in the deleted dimensions\footnote{Since the maximum entropy distribution under covariance constraint is a Gaussian.}, specifically, note that
\begin{align*}
h(X + Z) - h(\Pi_{\hat{\mathbf{V}}_d} X) &\geq \frac{D-d}{2} \log (2 \pi e \sigma^2),\\
\frac{D-d}{2} \log (2 \pi e &(\lambda_{d+1} + \sigma^2)) \geq \\
h(X + Z) - h(\Pi_{{\mathbf{V}}_d} X)  &\geq  \frac{D-d}{2} \log (2 \pi e \sigma^2).
\end{align*}
Hence, for small $d+1$th eigenvalue $\lambda_{d+1}$ and large enough samples, this dimensionality correction term $\frac{D-d}{2} \log (2 \pi e \sigma^2)$ will yield a tight approximation.

This procedure yields our PCA dimensionality reduced smoothed entropy estimator:
\begin{equation}\label{eq:PCAEstimator}
\hat{h}_{PCA}(X + Z) =  \hat{h}_{\sigma}(\Pi_{\hat{\mathbf{V}}_d} \mathbf{X}^{(2)}_n) + \frac{D-d}{2} \log (2 \pi e \sigma^2),
\end{equation}
where $\hat{h}_{\sigma}(\cdot)$ is the $d$-dimensional smoothed entropy estimator of \cite{goldfeld2019estimating}, specifically the plug-in estimator
\begin{equation}\label{eq:smoothEstimator}
\hat{h}_{\sigma}(\mathbf{X}^{(2)}_n) = h(P_{\mathbf{X}^{(2)}_n} \ast \mathcal{N}_\sigma)
\end{equation}
where $P_{\mathbf{X}^{(2)}_n}$ is the $d$-dimensional empirical distribution corresponding to the $n$ columns of $\mathbf{X}^{(2)}_n$ and $\mathcal{N}_\sigma$ is the distribution $\mathcal{N}(0, \sigma^2 I_d)$. As shown in \cite{goldfeld2019estimating}, this last entropy $h(P_{\mathbf{X}^{(2)}_n} \ast \mathcal{N}_\sigma)$ is simple to compute. 


Appendix \ref{app:MIest} applies our estimator to the noisy mutual information cases presented in the introduction.

\section{Convergence Rates for Low-Intrinsic Dimension Distributions}
We here derive bounds on the smoothed estimation error for a class of approximately low-dimensional data distributions. Our focus will be on approximately $d$-dimensional $X\in \mathbb{R}^D \sim P$ such that $d$-dimensional PCA recovers most of the variance of $X$. 

We consider the setting of sub-Gaussian $X$:\footnote{Slightly better constants can be achieved for bounded support densities, see \cite{goldfeld2020convergence} for details.}
\begin{definition}[sub-Gaussian Distribution]\label{DEF:SG}
A $d$-dimensional distribution $P$ is $K$-sub-Gaussian, for $K>0$, if $X\sim P$ satisfies
\begin{align}
\mathbb{E}&\left[\exp\left(\alpha^T(X \mspace{-2mu}-\mspace{-2mu} \mathbb{E} X)\right)\right] \leq \exp\big(0.5 K^2 \|\alpha\|^2\big),\nonumber\\ &\forall\alpha \in \mathbb{R}^d.\label{EQ:SG}
\end{align}
\end{definition}
In words, the above requires that every one-dimensional projection of $X$ be sub-Gaussian in the traditional scalar sense. When $\big(X-\mathbb{E} X\big)\in\mathcal{B}(0,R):=\big\{x\in\mathbb{R}^d\big|\|x\|_2\leq R\}$, \eqref{EQ:SG} holds with $K=R$.







\begin{theorem}\label{thm:main} Let $d$ be the target dimension of PCA, and $X$ a $K$-sub-Gaussian centered random vector in $\mathbb{R}^D$ such that 
\[
\mathbb{E}\|X\|_2^2\le M.
\] 
With $\lambda_1>\cdots>\lambda_D$ the eigenvalues of covariance $\Sigma$, we assume a $d$ to $d+1$ eigenvalue gap $\delta_d=\frac12(\lambda_d-\lambda_{d+1})$ and a residual eigenvalue sum 
${\sum_{i=d+1}^D\lambda_i}\le L$. Then for $c$ a constant,
\begin{align*}
&\mathbb{E}\abs{\hat{h}_{PCA}(X + Z)- h(X + Z)}\le \\&\frac{\log e}{\sigma^2}\p{3\sqrt{D\sigma^2+M}+4\sqrt{M}}\p{\sqrt{L}+\frac{2M^{3/2}}{\delta_d}n^{-1/2}}\\&+O_{\sigma,K}(c^d)n^{-1/2}.
\end{align*}
\end{theorem}
Note that the bound in Theorem \ref{thm:main} consists of two terms. The first term arises from errors in the PCA dimensionality reduction, and necessarily includes the residual eigenvalue sum $L$ as a persistent error.
The second term arises from the finite sample error of the reduced $d$-dimensional smoothed entropy estimator.
Importantly, note that the scaling with ambient dimension $D$ has been greatly reduced, instead of the $c^D$ scaling of the ambient space estimator, our bound scales only as $\sqrt{D}$ in the first term\footnote{The $\sqrt{D}$ is from the expected 2-norm of the noise $\mathbb{E}\|Z\|_2 = \sigma^2 D$, used in the Wasserstein continuity result of \cite{polyanskiy2016wasserstein}.} and $c^d$ in the second term as expected. Additionally, the scaling of the terms with $n$ remains fast at $O(n^{-1/2})$.

The proof of Theorem \ref{thm:main} given in Appendix \ref{app:proof}. The key building blocks of the proof are the Wasserstein continuity of smoothed entropy \cite{polyanskiy2016wasserstein} and finite sample bounds on the recovery of $d$-dimensional PCA projection matrices~\cite{DBLP:conf/nips/ZwaldB05}. 
The proof of the bound on the first term in Theorem \ref{thm:main} proceeds by (1) bounding the Frobenius norm error between the estimated projection matrix $\hat{\mathbf{V}}_d \hat{\mathbf{V}}_d^T$ and the true matrix ${\mathbf{V}}_d {\mathbf{V}}_d^T$, and (2) bounding the combined impact of the deleted residual terms $L$ and the error in estimating ${\mathbf{V}}_d$ on the projected smoothed entropy estimate using Wasserstein continuity and bounds on the associated Wasserstein distances. The bound on the second term follows from the $d$-dimensional smoothed entropy estimation result given in Theorem 3 of \cite{goldfeld2020convergence}.



\section{Empirical Results}

\subsection{Synthetic data entropy estimation}
\begin{figure}[h]
     \centering
     \begin{subfigure}{0.48\columnwidth}
       \centering
      \includegraphics[width=\columnwidth]{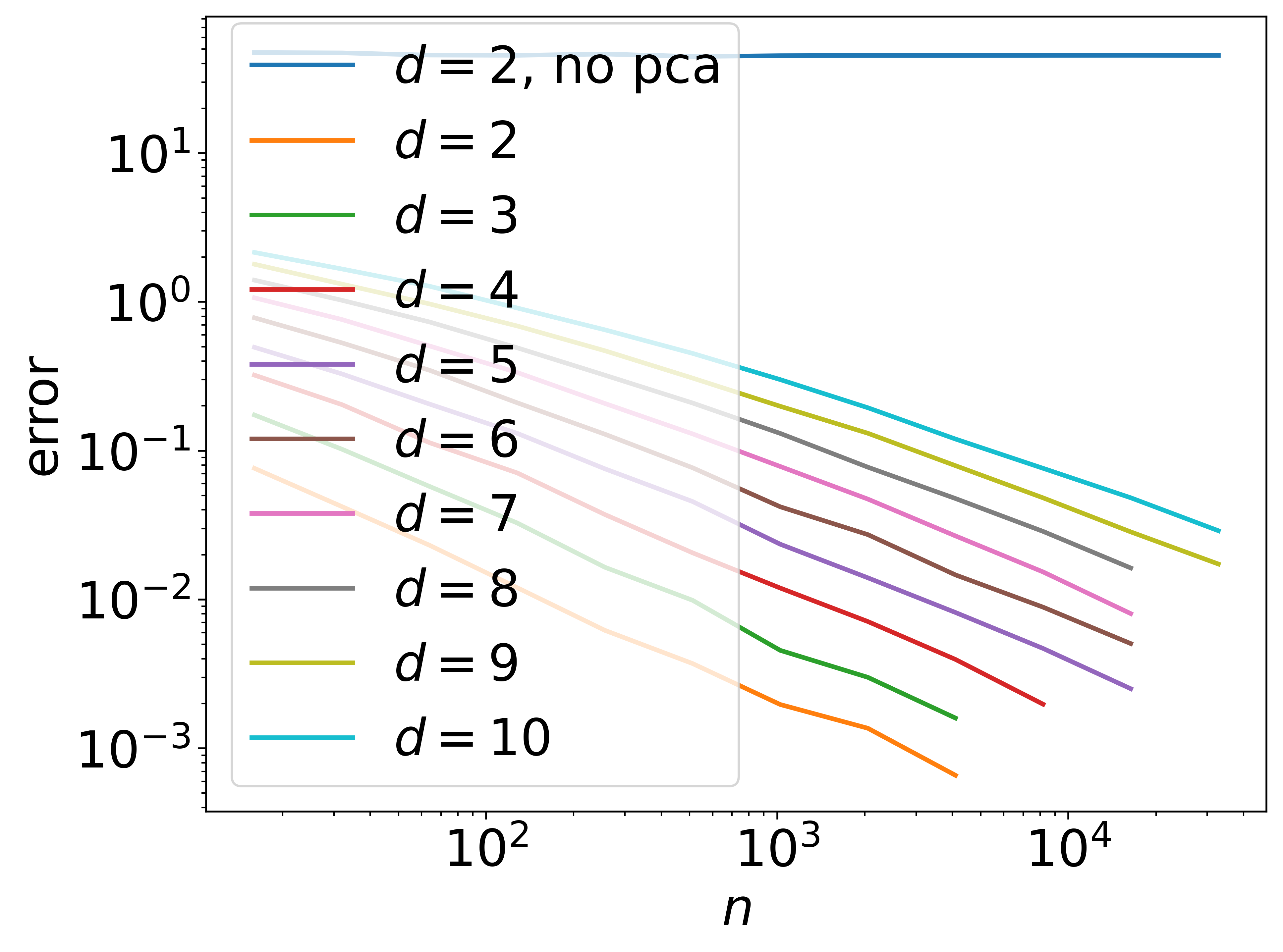}
      \caption{$2\le d\le 10,\,\sigma=.1, \,\lambda_D=.01$}\label{multi_d}
     \end{subfigure}%
     \hspace{.07in}
     \begin{subfigure}{0.48\columnwidth}
       \centering
      \includegraphics[width=\columnwidth]{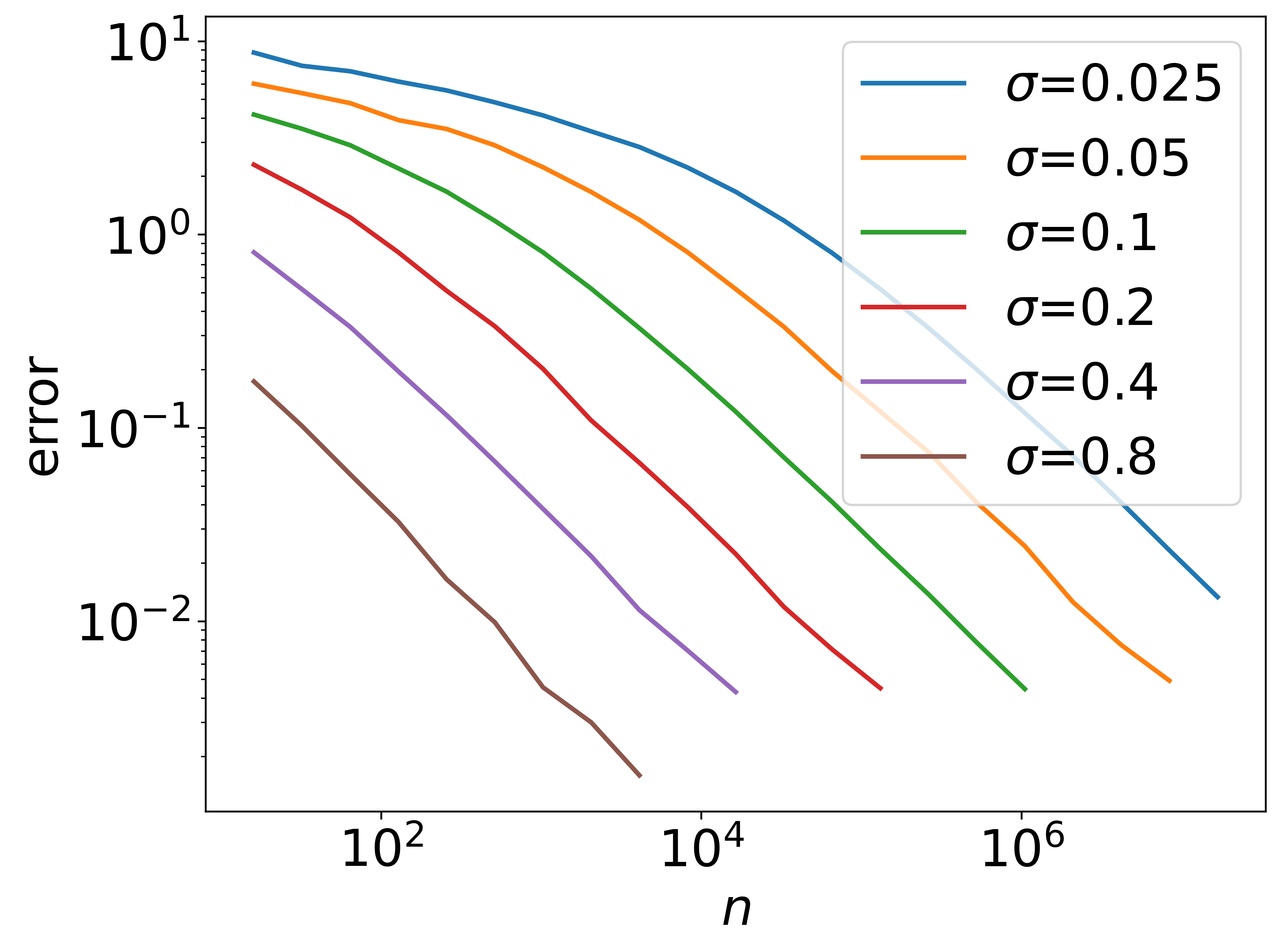}
      \caption{$.025\le \sigma\le .8,\,d=3,\,\lambda_D=.01$}\label{multiSigma}
     \end{subfigure}
     
     \begin{subfigure}{\columnwidth}
       \centering
      \includegraphics[width=.48\columnwidth]{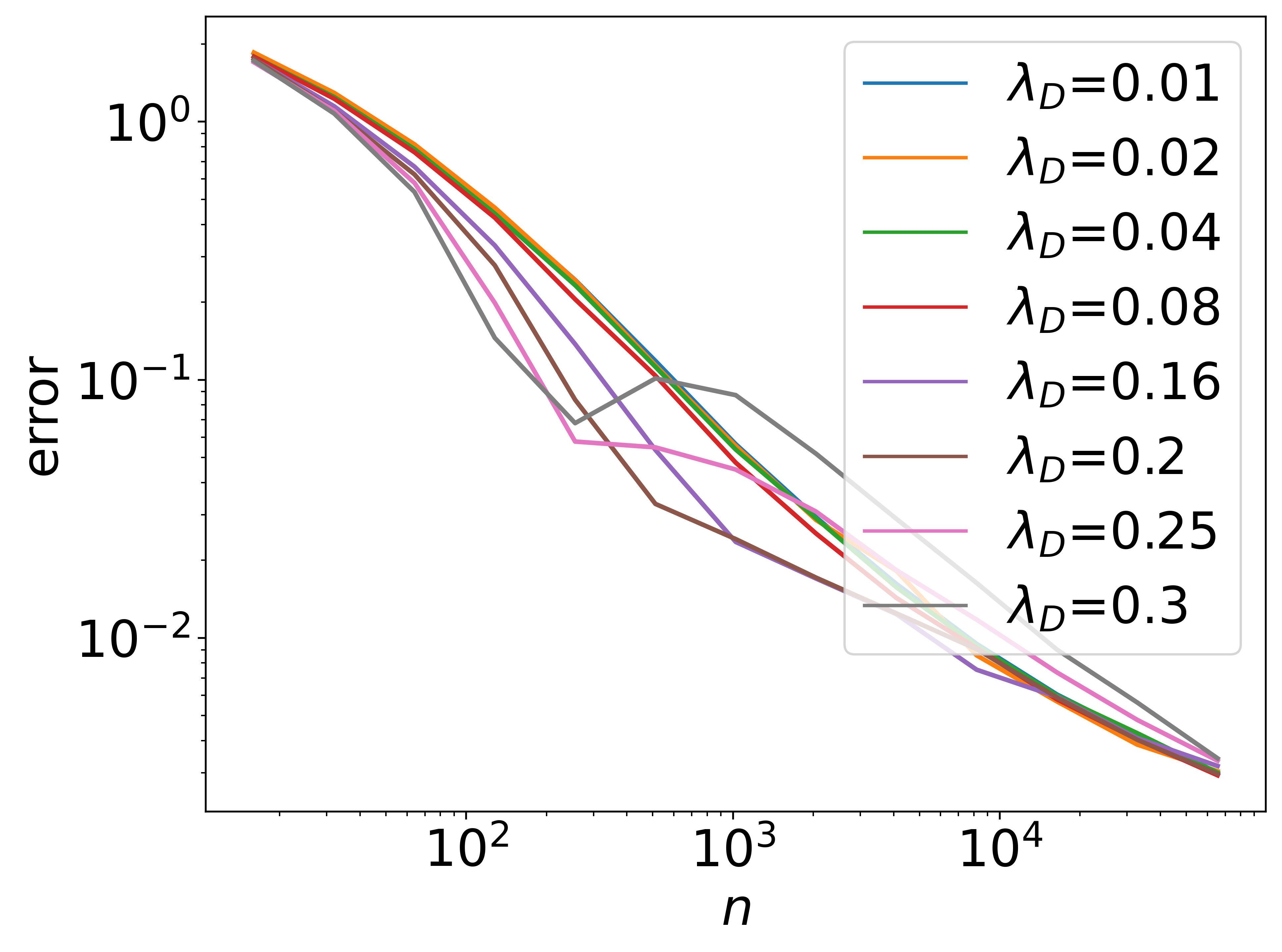}
      \caption{$.01\le \lambda_D\le .3,\,d=3,\,\sigma=.1$}\label{multiLambda}
     \end{subfigure}%
     \caption{Convergence of the (log-scale) error of estimating the $\sigma$-smoothed entropy of a $d$-dimensional Gaussian embedded in 100 dimensional space. Shown are parameter sweeps over number of samples $n$, dimension $d$, noise level $\sigma$, and intensity $\lambda_D$ of the $100-d$ residual eigenvalues. }\label{Gaussian}
     \vspace{-.2in}
    \end{figure}

We first verify the theoretical convergence result on synthetic data. 
We generated centered Gaussian $X$ with covariance $\Diag(\underbrace{1,\dots,1}_{d},\underbrace{\lambda_D,\dots,\lambda_D}_{100-d})$ and compared the post-PCA smoothed entropy estimate of $X+Z$ with the value of the ground-truth projected smoothed entropy, where $Z\sim\mathcal{N}(0,\sigma I_{100})$. Fig~\ref{Gaussian} shows that for varying $d,\sigma$, and $\lambda_D$, rapid convergence is observed despite the 100-dimensional ambient space. Note that higher $d$ and smaller $\sigma$ require more samples as expected, while the residual eigenvalues $\lambda_D$ do not have much effect on the error. The limited impact of $\lambda_D$ is to be expected while $\lambda_D$ remains somewhat smaller than 1, since PCA will successfully project away these eigenvectors in that regime, even though the sum of the $100-d$ eigenvalue intensities may be large.

\subsection{Synthetic data independence testing}
\begin{figure}[h]
\centering\includegraphics[width=.7\columnwidth]{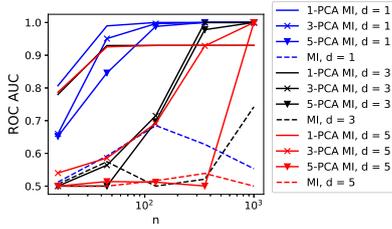}
\caption{Independence testing in $D = 100$ dimensional space, area under the ROC curve shown for thresholding estimated $\mathsf{I}(X + Z_1;Y+Z_2)$.}
\label{fig:indep}
\vspace{-.2in}
\end{figure}

We consider $X\in\mathbb{R}^{100}$ and $Y\in \mathbb{R}^{100}$ with a rank-$d$ common signal, specifically, $X$ and $Y$ being random projections of the same $d$-dimensional random Gaussian vector $W \sim \mathcal{N}(0,I_d)$ into $D=100$ dimensional space, i.e. $X = P_X W + N_X$, $Y = P_Y W + N_Y$, where $P_X, P_Y$ are $D\times d$ standard normal matrices, and $N_X, N_Y$ are independent Gaussian noise vectors with standard deviation 0.01. 
Figure \ref{fig:indep} shows independence testing results, where we estimate and threshold $\mathsf{I}(X + Z_1;Y+Z_2)$ (as in \eqref{eq:smoothjointMI}) using our PCA-based smoothed entropy estimator\footnote{This smoothed approach is equivalent to estimating $\mathsf{I}(X;Y)$ with a kernel density estimator.}, using $\sigma = 1$. Our dimensionality-reduced approach significantly outperforms ambient space mutual information estimation.

\subsection{Information flow in neural networks}

In this section, we address the problem of estimating information flow in noisy neural networks (where Gaussian noise of standard deviation $\sigma$ is added to each layer), as presented in \cite{goldfeld2019estimating}. We will show that the dimensionality reduction of PCA allows us to accurately track information flow with fewer samples than possible in the ambient space. We first test the fully connected neural network analyzed in \cite{DNNs_Tishby2017,goldfeld2019estimating}, and then use our approach to obtain similar plots in the much higher dimension regime of a convolutional neural network trained on the MNIST image dataset.

\begin{figure}[h]
     \centering
      \includegraphics[width=\columnwidth]{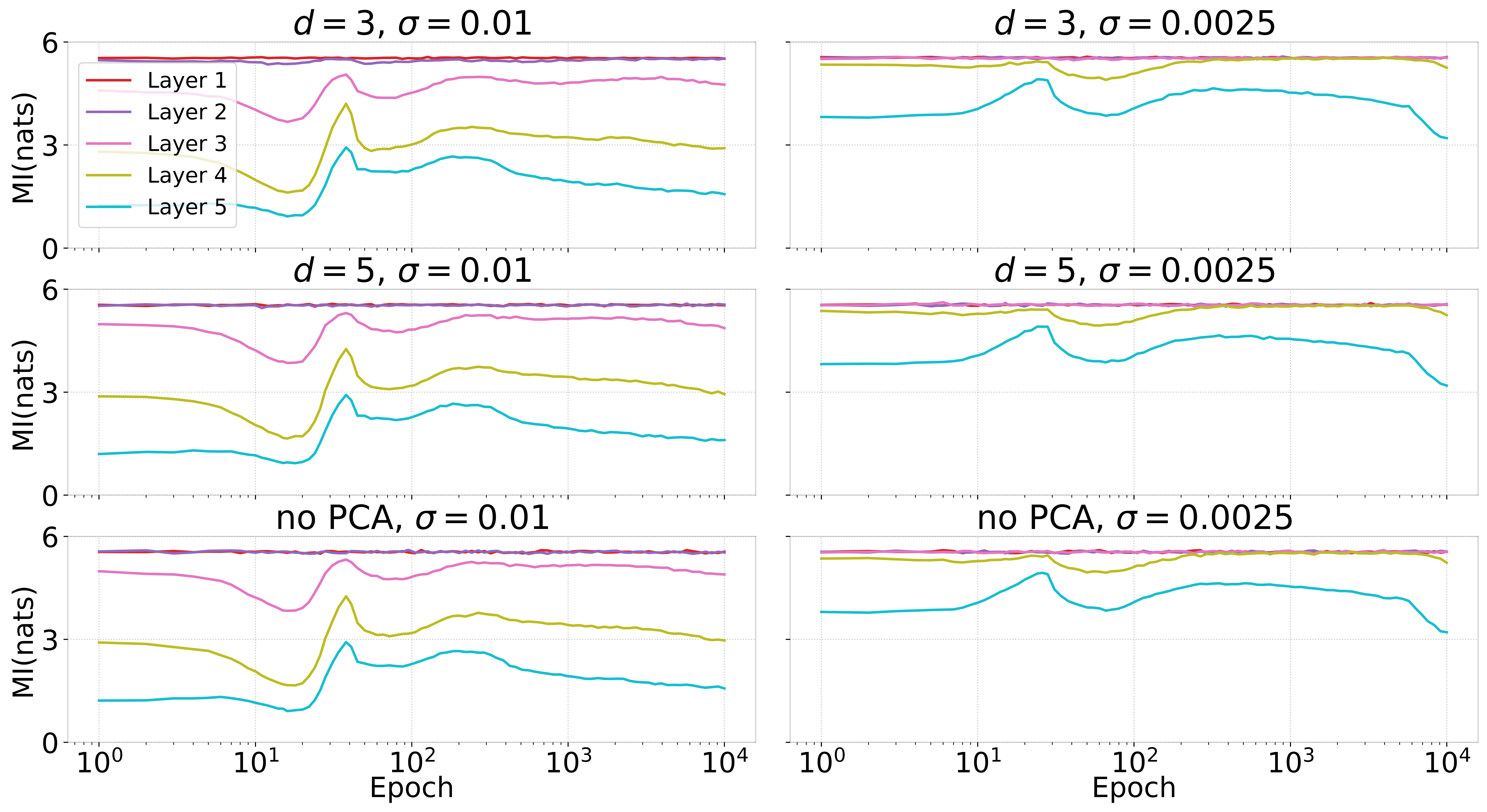}
     \caption{Evolution of MI across training epochs for the setting of~\cite{DNNs_Tishby2017} across PCA dimension $d$ and noise $\sigma$.}
             \label{fig:Tishby}
    \vspace{-.2in}
    \end{figure}

\paragraph{Fully connected neural network}
Consider the data and model of~\cite{DNNs_Tishby2017} for binary classification of 12-dimensional inputs. The noisy fully connected network is described in Appendix \ref{app:arch}. Results tracking estimates of the mutual information between the input $X$ and hidden layers $T_\ell$, for $\ell = 1,\dots,5$ are shown as a function of training epoch in Figure \ref{fig:Tishby}. Results are shown for PCA dimensions 3 and 5, along with no PCA, for small noise $\sigma= 0.0025$ and moderate noise $\sigma = 0.01$. Here we have 256 samples from $X$, and 40 samples of $T_\ell$ conditioned on each of the 256 values of $X$ for a total of $40 \times 256$ samples from $T_\ell$. Note that the $d=3,5$ PCA results successfully recover the trends shown in the ambient space estimates.

\begin{figure}[h!]
     \centering
      \includegraphics[width=\columnwidth]{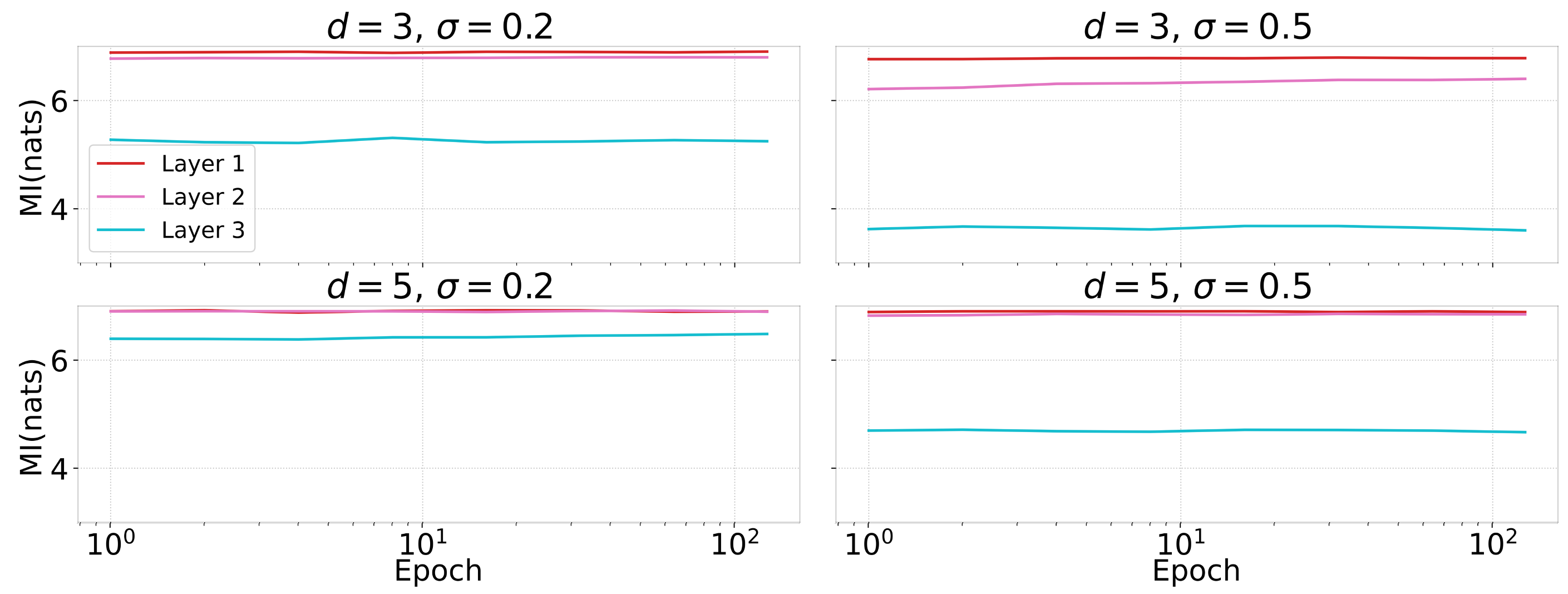}
     \caption{Evolution of MI across training epochs for the MNIST CNN across PCA dimension $d$ and noise $\sigma$.}
     \label{MNISTfig}
     \vspace{-.2in}
\end{figure}

\paragraph{Convolutional neural network for MNIST}
We trained a noisy convolutional neural network (CNN) on the MNIST dataset with architecture following \cite{goldfeld2019estimating} described in Appendix \ref{app:arch}. In contrast to the previous fully connected experiment, obtaining accurate smoothed entropy results was not practical computationally in \cite{goldfeld2019estimating} due to the needed sample complexity to estimate smoothed entropy in the high dimensional ambient space. Our dimensionality reduction approach indirectly reduces the computational requirements dramatically by reducing the sample complexity, allowing us to present converged (in terms of numbers of samples) estimates in that regime for PCA dimension $3$ and $5$ in Figure \ref{MNISTfig}. We used $N=10000$ samples.

Note that the mutual information estimates, particularly for the deeper layers, are not saturated (i.e. they are significantly smaller than the maximum value of $\ln N \approx 9.2$ where $N = 1000$ is the number of samples). Specifically, this indicates that the reduction in dimensionality has ensured that the limited-sample dataset is no longer mapped to the hidden layers injectively, as observed and discussed further in \cite{goldfeld2019estimating}).
Additionally, the mutual information estimates follow the data processing inequality, which need not occur in general if the PCA dimensionality reduction is not capturing enough of the variance. Regarding the information bottleneck interpretation of information flow, there does not seem to be much variation in the mutual information observed as a function of epoch. Investigating reasons for this behavior is an intriguing direction for future work using our methodology.




\section{Conclusion}
We addressed the problem of estimating smoothed entropy in high dimensions when the underlying distribution is approximately low dimensional. Using the structure provided by smoothing, we derived strong approximation and estimation guarantees in the low dimensional regime, demonstrating removal of the curse of dimensionality. We applied our approach to synthetic data experiments as well as an application to tracking information flow in neural networks.


\bibliographystyle{IEEEtran}
\bibliography{PCAMI}

\begin{thebibliography}{10}
\providecommand{\url}[1]{#1}
\csname url@samestyle\endcsname
\providecommand{\newblock}{\relax}
\providecommand{\bibinfo}[2]{#2}
\providecommand{\BIBentrySTDinterwordspacing}{\spaceskip=0pt\relax}
\providecommand{\BIBentryALTinterwordstretchfactor}{4}
\providecommand{\BIBentryALTinterwordspacing}{\spaceskip=\fontdimen2\font plus
\BIBentryALTinterwordstretchfactor\fontdimen3\font minus
  \fontdimen4\font\relax}
\providecommand{\BIBforeignlanguage}[2]{{%
\expandafter\ifx\csname l@#1\endcsname\relax
\typeout{** WARNING: IEEEtran.bst: No hyphenation pattern has been}%
\typeout{** loaded for the language `#1'. Using the pattern for}%
\typeout{** the default language instead.}%
\else
\language=\csname l@#1\endcsname
\fi
#2}}
\providecommand{\BIBdecl}{\relax}
\BIBdecl

\bibitem{goldfeld2019estimating}
Z.~Goldfeld, E.~van~den Berg, K.~H. Greenewald, I.~Melnyk, N.~Nguyen,
  B.~Kingsbury, and Y.~Polyanskiy, ``Estimating information flow in deep neural
  networks,'' in \emph{ICML}, 2019.

\bibitem{goldfeld2020convergence}
Z.~Goldfeld, K.~Greenewald, J.~Niles-Weed, and Y.~Polyanskiy, ``Convergence of
  smoothed empirical measures with applications to entropy estimation,''
  \emph{IEEE Transactions on Information Theory}, vol.~66, no.~7, pp.
  4368--4391, 2020.

\bibitem{tishby_DNN2015}
N.~Tishby and N.~Zaslavsky, ``Deep learning and the information bottleneck
  principle,'' in \emph{Proceedings of the Information Theory Workshop (ITW)},
  Jerusalem, Israel, Apr.-May 2015, pp. 1--5.

\bibitem{xu2017information}
A.~Xu and M.~Raginsky, ``Information-theoretic analysis of generalization
  capability of learning algorithms,'' \emph{Advances in Neural Information
  Processing Systems}, vol.~30, 2017.

\bibitem{steinke2020reasoning}
T.~Steinke and L.~Zakynthinou, ``Reasoning about generalization via conditional
  mutual information,'' in \emph{Conference on Learning Theory}.\hskip 1em plus
  0.5em minus 0.4em\relax PMLR, 2020, pp. 3437--3452.

\bibitem{berrett2019nonparametric}
T.~B. Berrett and R.~J. Samworth, ``Nonparametric independence testing via
  mutual information,'' \emph{Biometrika}, vol. 106, no.~3, pp. 547--566, 2019.

\bibitem{sen2017model}
R.~Sen, A.~T. Suresh, K.~Shanmugam, A.~G. Dimakis, and S.~Shakkottai,
  ``Model-powered conditional independence test,'' \emph{Advances in neural
  information processing systems}, vol.~30, 2017.

\bibitem{belghazi2018mine}
M.~I. Belghazi, A.~Baratin, S.~Rajeswar, S.~Ozair, Y.~Bengio, A.~Courville, and
  R.~D. Hjelm, ``Mine: mutual information neural estimation,'' \emph{arXiv
  preprint arXiv:1801.04062}, 2018.

\bibitem{chen2016infogan}
X.~Chen, Y.~Duan, R.~Houthooft, J.~Schulman, I.~Sutskever, and P.~Abbeel,
  ``Infogan: Interpretable representation learning by information maximizing
  generative adversarial nets,'' \emph{Advances in neural information
  processing systems}, vol.~29, 2016.

\bibitem{DNNs_ICLR2018}
A.~M. Saxe, Y.~Bansal, J.~Dapello, M.~Advani, A.~Kolchinsky, B.~D. Tracey, and
  D.~D. Cox, ``On the information bottleneck theory of deep learning,'' in
  \emph{Proceedings of the International Conference on Learning Representations
  (ICLR)}, 2018.

\bibitem{jacobsen2018revnet}
J.-H. Jacobsen, A.~Smeulders, and E.~Oyallon, ``{i-RevNet}: Deep invertible
  networks,'' \emph{arXiv preprint arXiv:1802.07088}, 2018.

\bibitem{liu2018understanding}
K.~Liu, R.~A. Amjad, and B.~C. Geiger, ``Understanding individual neuron
  importance using information theory,'' \emph{arXiv preprint
  arXiv:1804.06679}, 2018.

\bibitem{gabrie2018entropy}
M.~Gabri{\'e}, A.~Manoel, C.~Luneau, J.~Barbier, N.~Macris, F.~Krzakala, and
  L.~Zdeborov{\'a}, ``Entropy and mutual information in models of deep neural
  networks,'' \emph{arXiv preprint arXiv:1805.09785}, 2018.

\bibitem{reeves2017additivity}
G.~Reeves, ``Additivity of information in multilayer networks via additive
  gaussian noise transforms,'' in \emph{Proc. 55th Annu. Allerton Conf.
  Commun., Control and Comput. (Allerton-2017)}, Monticello, Illinois, Oct.
  2017, pp. 1064--1070.

\bibitem{DNNs_Tishby2017}
R.~Shwartz-Ziv and N.~Tishby, ``Opening the black box of deep neural networks
  via information,'' 2017, arXiv:1703.00810 [cs.LG].

\bibitem{bengio2013representation}
Y.~Bengio, A.~Courville, and P.~Vincent, ``Representation learning: A review
  and new perspectives,'' \emph{IEEE transactions on pattern analysis and
  machine intelligence}, vol.~35, no.~8, pp. 1798--1828, 2013.

\bibitem{jiao2017nearest}
J.~Jiao, W.~Gao, and Y.~Han, ``The nearest neighbor information estimator is
  adaptively near minimax rate-optimal,'' \emph{arXiv preprint
  arXiv:1711.08824}, 2017.

\bibitem{Hero_EDGE2018}
M.~Noshad and A.~O.~H. III, ``Scalable mutual information estimation using
  dependence graphs,'' 2018, arXiv:1801.09125 [cs.IT].

\bibitem{poole2019variational}
B.~Poole, S.~Ozair, A.~Van Den~Oord, A.~Alemi, and G.~Tucker, ``On variational
  bounds of mutual information,'' in \emph{International Conference on Machine
  Learning}.\hskip 1em plus 0.5em minus 0.4em\relax PMLR, 2019, pp. 5171--5180.

\bibitem{goldfeld2022k}
Z.~Goldfeld, K.~Greenewald, T.~Nuradha, and G.~Reeves, ``k-sliced mutual
  information: A quantitative study of scalability with dimension,'' in
  \emph{Annual Conference on Neural Information Processing Systems}, 2022.

\bibitem{balasubramanian2002isomap}
M.~Balasubramanian and E.~L. Schwartz, ``The isomap algorithm and topological
  stability,'' \emph{Science}, vol. 295, no. 5552, pp. 7--7, 2002.

\bibitem{zbontar2021barlow}
J.~Zbontar, L.~Jing, I.~Misra, Y.~LeCun, and S.~Deny, ``Barlow twins:
  Self-supervised learning via redundancy reduction,'' in \emph{International
  Conference on Machine Learning}.\hskip 1em plus 0.5em minus 0.4em\relax PMLR,
  2021, pp. 12\,310--12\,320.

\bibitem{polyanskiy2016wasserstein}
Y.~Polyanskiy and Y.~Wu, ``Wasserstein continuity of entropy and outer bounds
  for interference channels,'' \emph{IEEE Transactions on Information Theory},
  vol.~62, no.~7, pp. 3992--4002, 2016.

\bibitem{DBLP:conf/nips/ZwaldB05}
\BIBentryALTinterwordspacing
L.~Zwald and G.~Blanchard, ``On the convergence of eigenspaces in kernel
  principal component analysis,'' in \emph{Advances in Neural Information
  Processing Systems 18 [Neural Information Processing Systems, {NIPS} 2005,
  December 5-8, 2005, Vancouver, British Columbia, Canada]}, 2005, pp.
  1649--1656. [Online]. Available:
  \url{https://proceedings.neurips.cc/paper/2005/hash/2b6921f2c64dee16ba21ebf17f3c2c92-Abstract.html}
\BIBentrySTDinterwordspacing

\end{thebibliography}
\flushcolsend
\clearpage

\appendices
\section{Computation of the Smoothed Entropy Estimator}
\label{app:computation}
Evaluating the plug-in estimator $\hat{h}_\sigma(\mathbf{X}'_n) = h(\hat{P}_{\mathbf{X}_n}\ast\mathcal{N}_\sigma)$ requires computing the differential entropy of the $d$-dimensional $n$-mode Gaussian mixture ($\hat{P}_{\mathbf{X}_n}\ast\mathcal{N}_\sigma$). While this cannot be computed in closed form, \cite{goldfeld2020convergence} proposed the following computationally efficient Monte Carlo (MC) integration based approach, whose MC error is bounded theoretically therein. We describe this procedure for completeness.


Let $g(t):=\hat{P}_{\mathbf{X}_n}\ast\mathcal{N}_\sigma = \frac{1}{n}\sum_{i=1}^n \phi_\sigma(t-X_i)$, where $\phi_\sigma$ is the density of $\mathcal{N}_\sigma$. Let $C\sim\mathsf{Unif}\p{\{X_i\}_{i=1}^n}$ be independent of $Z\sim\mathcal{N}_\sigma$ and note that $V:= C+Z\sim g$. We can write $h(g)$ as:
\begin{align}
    h(g)=-\mathbb{E}\log g(V)&=-\frac{1}{n}\sum_{i=1}^n\mathbb{E}\Big[\log g(X_i+Z)\Big|C=\mu_i\Big]\nonumber\\&=-\frac{1}{n}\sum_{i=1}^n\mathbb{E}\log g(X_i+Z).\label{EQ:MC_expansion}
\end{align}
Let $\left\{Z_j^{(i)}\right\}_{\substack{i\in[n]\\j\in[n_{\mathsf{MC}}]}}$ be $n\times n_{\mathsf{MC}}$ i.i.d. samples from $\mathcal{N}_\sigma$, where $n_{\mathsf{MC}}$ is the chosen number of Monte Carlo trials. For each $i\in[n]$, we estimate the $i$-th summand on the RHS of \eqref{EQ:MC_expansion} by
\begin{subequations}
\begin{equation}
    \hat{{I}}_\mathsf{MC}^{(i)}:= \frac{1}{n_{\mathsf{MC}}}\sum_{j=1}^{n_{\mathsf{MC}}}\log g\left(\mu_i+Z_j^{(i)}\right),\label{EQ:MC_per_i_est}
\end{equation}
which produces 
\begin{equation}
    \hat{h}_\mathsf{MC}:= \frac{1}{n}\sum_{i=1}^n\hat{I}_\mathsf{MC}^{(i)}\label{EQ:MC_full_est}
\end{equation}\label{EQ:MCI}%
\end{subequations}
as the approximation of $h(g)$. Note that since $g$ is a mixture of $n$ Gaussians, it can be efficiently evaluated using off-the-shelf KDE software packages, many of which require only $O(\log n)$ operations on average per evaluation of $g$. 

\section{Proof of Theorem \ref{thm:main}}
\label{app:proof}
\begin{IEEEproof} 
Recall $\hat{\mathbf{V}}_d$ and ${\mathbf{V}}_d$ are $D \times d$ matrices of the top $d$ eigenvectors of $\Sigma$ and $\Sigma_n$ respectively. Note that $\hat{\mathbf{V}}^T_d X$ projects $X$ to $d$-dimensional space, and $\left(\hat{\mathbf{V}}_d\hat{\mathbf{V}}^T_d\right) X$ projects $X$ to a $d$-dimensional hyperplane, still contained in the ambient $\mathbb{R}^D$ space.

We make use of the 2-Wasserstein distance, where the squared 2-Wasserstein distance between measures $\mu$ and $\nu$ is $ \mathsf{W}^2_2(\mu,\nu) := \inf \E\|X - Y\|^2$, with the infimum taken over all couplings of $\mu$ and $\nu$.

We will decompose the entropy estimation error into error arising from two sources: (a) approximation of $X$ by a PCA-based projection estimated using the $n$ samples $\mathbf{X}^{(1)}_n$, and (b) estimation error of the smoothed entropy estimator in the resulting $d$-dimensional projected space, using the held out $n$ samples $\mathbf{X}^{(2)}_n$.

\paragraph{Error of approximating the ambient entropy by PCA-estimated projected entropy} First, we relate the smoothed entropy of $X$ to the smoothed entropy of $X$ after projection to the $d$-dimensional empirical PCA (based on the empirical sample covariance $\Sigma_n$) hyperplane, specifically, the smoothed entropy of $\hat W := \left(\hat{\mathbf{V}}_d\hat{\mathbf{V}}^T_d\right) X$. Similarly, let the projection to the $d$-dimensional oracle PCA (i.e. based on the true covariance $\Sigma$) be $W := \left({\mathbf{V}}_d{\mathbf{V}}^T_d\right) X$.

Using Corollary 4 of \cite{polyanskiy2016wasserstein}, we bound 
$\abs{h(X+Z) - h(W+Z)}$  in terms of the Wasserstein-2 distances $\mathsf{W}_2(X+Z,{W} + Z)$ and $\mathsf{W}_2({W} + Z,\hat W + Z)$:
\begin{align*}
&\abs{h(X + Z) - h(W + Z)}\\
&\le \frac{\log e}{\sigma^2}\p{3\sqrt{D\sigma^2+M}+4\sqrt{M}}\mathsf{W}_2(X+Z,\hat{W}+Z)\\
&\le\frac{\log e}{\sigma^2}\p{3\sqrt{D\sigma^2+M}+4\sqrt{M}}\\
&\quad\times\p{\mathsf{W}_2(X+Z,W + Z)+\mathsf{W}_2(W+Z,\hat{W} +Z)},
\end{align*}
where in the last step we have used the triangle inequality.

To bound the first distance $\mathsf{W}_2(X+Z, W+Z)$, we have
\begin{align*}
\mathsf{W}_2(X+Z,W+Z) 
&\le \mathsf{W}_2(X,W)\\
&\le\sqrt{\E{\norm{X-W}^2}}\\
&=\sqrt{{\sum_{i=d+1}^D\lambda_i}}\\
&\le \sqrt{L},
\end{align*}
where we have used first that the Wasserstein distance is non-increasing under convolutions and then bounded the optimally-coupled Wasserstein distance by the cost of coupling via the map $W = \left({\mathbf{V}}_d{\mathbf{V}}^T_d\right) X$ from the definition of $W$.

To bound $\mathsf{W}_2(W+Z,\hat{W} + Z)$, we first relate the oracle and estimated projection matrices.
By Lemma 1 of~\cite{DBLP:conf/nips/ZwaldB05},
\[
\E{\norm{\Sigma_n-\Sigma}^2_{F}}\le\frac{4M^2}{n},
\]
where $\|\cdot\|_F$ is the Frobenius norm.
Furthermore, by Theorem 3 of~\cite{DBLP:conf/nips/ZwaldB05},
\[
\E{\norm{\hat{\mathbf{V}}_d\hat{\mathbf{V}}^T_d-{\mathbf{V}}_d{\mathbf{V}}^T_d}^2_{F}}\le\frac{\E{\norm{\Sigma_n-\Sigma}^2_{F}}}{\delta_d^2}\le\frac{4M^2}{\delta_d^2n},
\]
where $\delta_d=\frac12(\lambda_d-\lambda_{d+1})$ is the gap between the $d$th and $d+1$th eigenvalue.\footnote{Intuitively, the larger this gap is, the easier it is to identify the $d$-dimensional subspace with largest variance. } Then
\begin{align*}
\mathsf{W}_2&(\hat W + Z,{W} + Z) 
\le \mathsf{W}_2\left(\left(\hat{\mathbf{V}}_d\hat{\mathbf{V}}^T_d\right)X,\left({\mathbf{V}}_d{\mathbf{V}}^T_d\right)X\right)\\
&\le\sqrt{\E{\left\|{\p{\hat{\mathbf{V}}_d\hat{\mathbf{V}}^T_d-{\mathbf{V}}_d{\mathbf{V}}^T_d}X}\right\|_2^2}}\\
&\le\sqrt{\E{\left[\left\|\hat{\mathbf{V}}_d\hat{\mathbf{V}}^T_d-{\mathbf{V}}_d{\mathbf{V}}^T_d\right\|_2^2\norm{X}_2^2\right]}}\\
&=\sqrt{\E{\norm{\hat{\mathbf{V}}_d\hat{\mathbf{V}}^T_d-{\mathbf{V}}_d{\mathbf{V}}^T_d}_2^2}\cdot\E{\norm{X}_2^2}}\\
&\le \sqrt{\E{\left\|\hat{\mathbf{V}}_d\hat{\mathbf{V}}^T_d-{\mathbf{V}}_d{\mathbf{V}}^T_d\right\|_{F}^2}\cdot M}\\
&\le\frac{2M^{3/2}}{\delta_d}n^{-1/2}.
\end{align*}
where for matrices, $\|\cdot\|_2$ denotes the spectral norm. The first inequality again follows since the Wasserstein distance is non-increasing under convolutions, and the remainder follow from the independence of $X$ and $\hat{\mathbf{V}}_d$ and the norm inequalities $\|A X\|_2 \leq \|A\|_2 \|X\|_2$, $\|A\|_2 \leq \|A\|_F$.

\paragraph{Error of the smoothed entropy estimator in the latent space} It remains to relate the smoothed entropy of $\hat{W}$ in the ambient space to the smoothed entropy of $\Pi_{\hat{\mathbf{V}}_d} X$ in the latent ($d$-dimensional) space. Noting that by definition
\[
\Pi_{\hat{\mathbf{V}}_d} (X+Z) = \Pi_{\hat{\mathbf{V}}_d} (\hat{W} + Z) 
\]
and $\Pi_{\hat{\mathbf{V}}_d} Z \sim \mathcal{N}(0,\sigma^2 I_d)$,
\[
h(\Pi_{\hat{\mathbf{V}}_d} (X + Z))= h(W + Z)-\frac{D-d}{2}\log(2\pi e\sigma^2).
\]

It remains to account for finite sample estimation error of this projected smoothed entropy using the samples $\mathbf{X}_n^{(2)}$. By Theorem 3 of \cite{goldfeld2020convergence},
\[
\mathbb{E}{\abs{\hat{h}_\sigma(\Pi_{\hat{\mathbf{V}}_d} \mathbf{X}^{(2)}_n)-h(\Pi_{\hat{\mathbf{V}}_d} (X + Z))}}\le O_{\sigma,K}(c^d)n^{-1/2},
\]
where $c$ is a constant.

Putting the above together yields the theorem.
\end{IEEEproof}

\section{Application to Mutual Information Estimation} 
\label{app:MIest}
We apply the smooth low-dimensional entropy estimator to the two noisy mutual information estimation cases (conditional vs. joint sampling) presented in \eqref{eq:smoothMI} and \eqref{eq:smoothjointMI} in the introduction respectively. 

\paragraph{Case 1: Conditional Sampling} In the case \eqref{eq:smoothMI}, the $h(Y+Z)$ term can be estimated via our estimator \eqref{eq:PCAEstimator} straightforwardly. The $h(Y + Z|X)$ term can be estimated using the expression
\[
h(Y + Z | X) = \mathbb{E}_{x \sim p_X} [h(Y + Z | X = x)] 
\]
and using $m$-trial Monte Carlo averaging as
\[
\hat{h}(Y+Z | X) = \sum_{i = 1}^m \hat{h}(Y + Z | X = x_i)
\]
where the $x_i \sim p_X$ and the $\hat{h}(Y + Z | X = x_i)$ estimates are computed using our estimator \eqref{eq:PCAEstimator} using $n$ samples from the conditional distribution $p_{Y | X = x_i}$.

\paragraph{Case 2: Joint Sampling} In the case \eqref{eq:smoothjointMI}, the $h(X+Z_1)$, $h(Y + Z_2)$ terms can be estimated straightforwardly using our $d$-dimensional estimator \eqref{eq:PCAEstimator}. The joint entropy term $h(X + Z_1, Y + Z_2)$ is now a $2D$-dimensional entropy, hence it is appropriate to use the estimator \eqref{eq:PCAEstimator} on the $2D$-dimensional vector 
\[
\left[\begin{array}{c} X \\ Y \end{array} \right] + Z
\]
where $Z \sim \mathcal{N}(0, \sigma^2 I_{2D})$ is $2D$-dimensional $\sigma$-Gaussian noise, and now the dimensionality is reduced to $2d$ instead of $d$.

\section{Neural Network Architectures}
\label{app:arch}
\paragraph{Fully connected neural network for SZT model} Following \cite{DNNs_Tishby2017}, our fully connected neural network architecture for binary classification has 12-dimensional inputs, with seven hidden layers with 12--10--7--5--4--3--2 hidden units respectively. We used tanh nonlinearities, with i.i.d. Gaussian noise of standard deviation $\beta$ injected in each layer as in \cite{goldfeld2019estimating}. 

\paragraph{Convolutional neural network for MNIST}
We used the following architecture first presented in \cite{goldfeld2019estimating}, including two convolutional layers, two fully connected layers, and batch normalization.
\begin{enumerate}
\item 2-d convolutional layer with 1 input channel, 16 output channels, 5x5 kernels, and input padding of 2 pixels
\item Batch normalization
\item Tanh() activation function
\item Zero-mean additive Gaussian noise with variance $\sigma$
\item 2x2 max-pooling
\item 2-d convolutional layer with 16 input channels, 32 output channels, 5x5 kernels, and input padding of 2 pixels
\item Batch normalization
\item Tanh() activation function
\item Zero-mean additive Gaussian noise with variance $\sigma$
\item 2x2 max-pooling
\item Fully connected layer with 1586 (32x7x7) inputs and 128 outputs
\item Batch normalization
\item Tanh() activation function
\item Zero-mean additive Gaussian noise with variance $\sigma$
\item Fully connected layer with 128 inputs and 10 outputs
\end{enumerate}
All convolutional and fully connected layers have weights and biases, and the weights are initialized using the default initialization, which draws weights from $\mathsf{Unif}[-1/\sqrt{m},1/\sqrt{m}]$, with $m$ the fan-in to a neuron in the layer. Training uses cross-entropy loss, and is performed using stochastic gradient descent with no momentum, 128 training epochs, and 32-sample minibatches. The initial learning rate is $5 \times 10^{-3}$, and it is reduced following a geometric schedule such that the learning rate in the final epoch is $5 \times 10^{-4}$. To improve the test set performance of our models, we applied data augmentation to the training set by translating, rotating, and shear-transforming each training example each time it was selected. Translations in the $x$- and $y$-directions were drawn uniformly from $\{-2,-1,0,1,2\}$, rotations were drawn from $\mathsf{Unif}(-10^\circ,10^\circ)$, and shear transforms were drawn from $\mathsf{Unif}(-10^\circ,10^\circ)$.

\section{Additional experiments}
\subsection{Spiral data embedded in high dimensional space}
\label{app:spiral}
\begin{figure}[h!]
     \centering
     \begin{subfigure}{.5\columnwidth}
       \centering
      \includegraphics[width=\columnwidth]{multiLambda.png}
      \caption{2-d spiral with $.01\le \lambda_D\le .3,\,\sigma=.1$}\label{multiLambda_2_helix}
     \end{subfigure}%
     \begin{subfigure}{.5\columnwidth}
       \centering
      \includegraphics[width=\columnwidth]{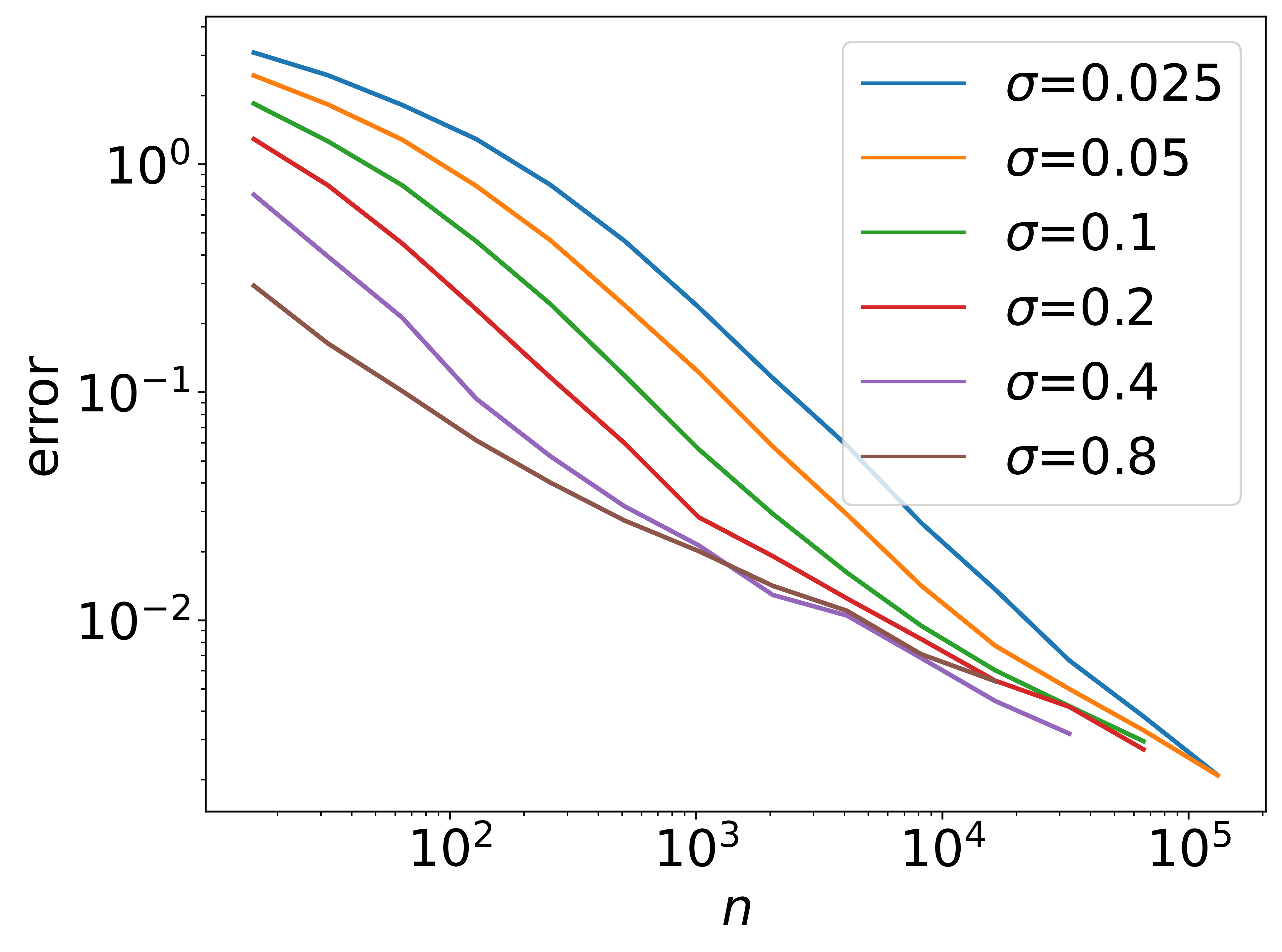}
      \caption{2-d spiral with $.025\le \sigma\le .8,\,\lambda_D=.01$}\label{multiSigma_2_helix}
     \end{subfigure}
     
     \vskip\baselineskip
     \begin{subfigure}{.5\columnwidth}
       \centering
      \includegraphics[width=\columnwidth]{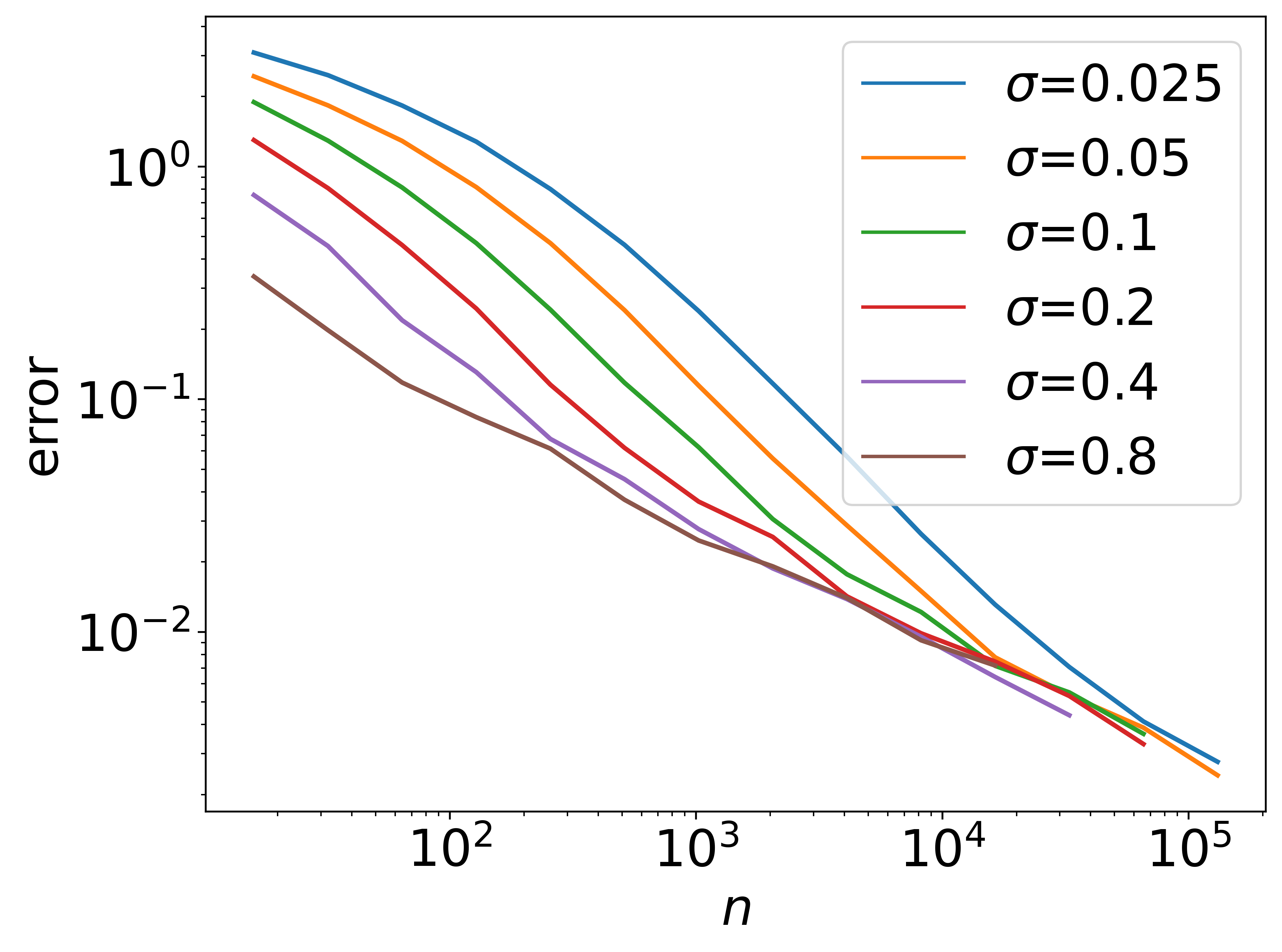}
      \caption{3-d conical spiral with $.025\le \sigma\le .8,\lambda_D=.01$}\label{multiLambda_2_helix}
     \end{subfigure}%
     \begin{subfigure}{.5\columnwidth}
       \centering
      \includegraphics[width=\columnwidth]{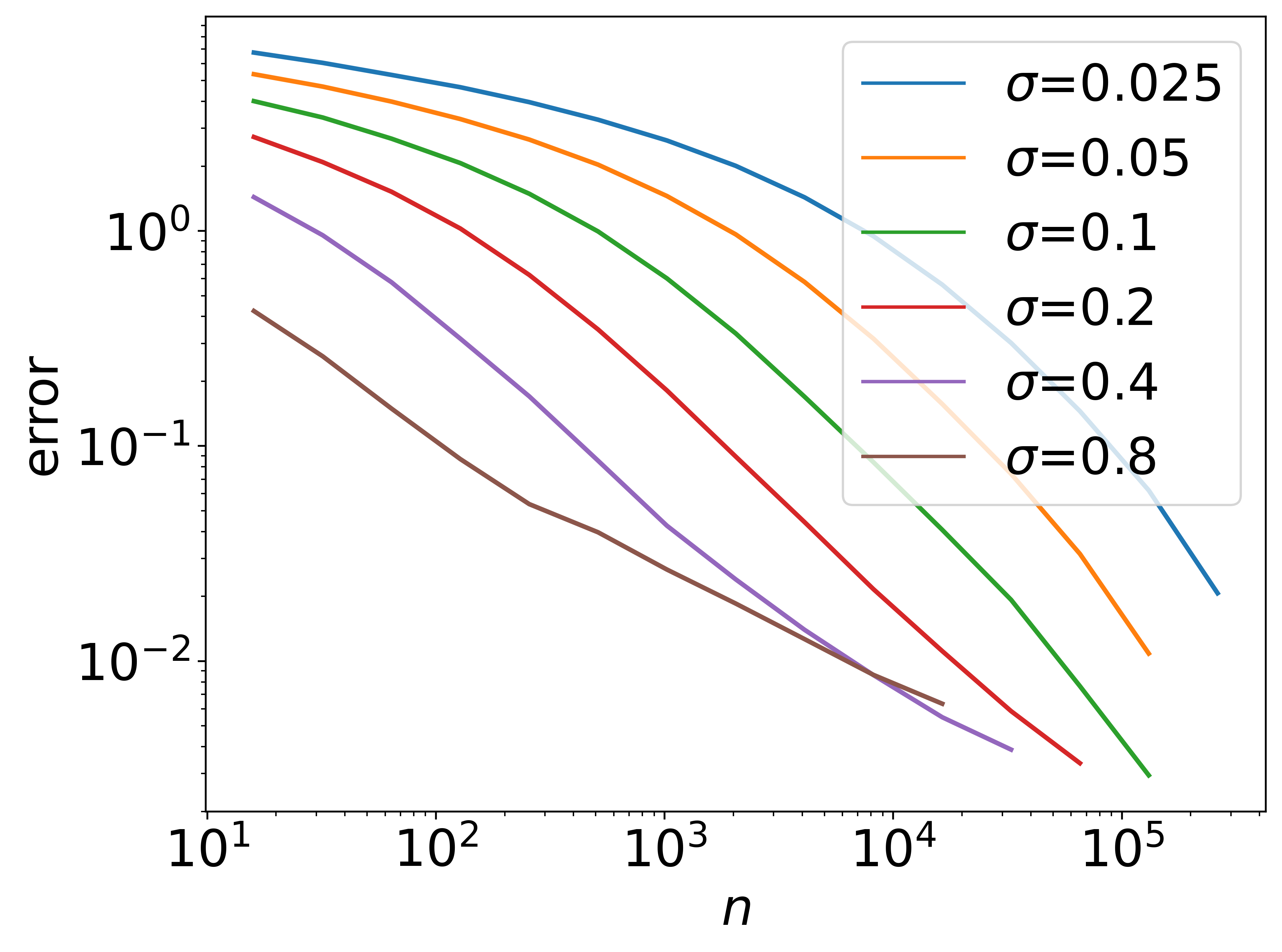}
      \caption{3-d cylindrical spiral with $.025\le \sigma\le .8,\lambda_D=.01$}\label{multiSigma_2_helix}
     \end{subfigure}
     \caption{Convergence of the (log-scale) error of estimating the $\sigma$-smoothed entropy of low dimensional spiral datasets embedded in 100-dimensional space.}\label{spiral}
    \end{figure}

We first generated three types of spiral data $S$: a 2-dimensional spiral given by $(r\cos\theta,r\sin\theta)$, and 3-dimensional conical and cylindrical spirals given by $(r\cos\theta,r\sin\theta,r)$ and $(r\cos\theta,r\sin\theta,z)$ respectively, where $z\sim\Unif([0,4])$. Then we generated 100-dimensional $X=(S,T)$, where $T\sim\mathcal{N}(0,\lambda_D^2 I_{100-d})$, and compared the post-PCA entropy estimate of $X+Z$ with the ground truth, where $Z\sim\mathcal{N}(0,\sigma I_{100})$. Fig~\ref{spiral} shows for varying spiral type, $\sigma$, and $\lambda_D$, rapid convergence is observed despite the 100-dimensional ambient space. Note that as with the embedded Gaussian setting, error is reduced by increasing $\sigma$, and $\lambda_D$ has limited impact. 
    \flushcolsend
\end{document}